# Revealing surface states in In-doped SnTe superconducting nanoplates with low bulk mobility

Jie Shen[1,2], Yujun Xie[1,2], Judy J. Cha[1,2*]

[1] Department of Mechanical Engineering and Materials Science, Yale University, New Haven, CT, USA

[2] Energy Sciences Institute, Yale West Campus, West Haven, CT, USA

**Corresponding Author**

* E-mail: judy.cha@yale.edu

ABSTRACT

Indium (In) doping in topological crystalline insulator SnTe induces superconductivity, making In-doped SnTe a candidate for a topological superconductor. SnTe nanostructures offer well-defined nanoscale morphology and high surface-to-volume ratios to enhance surface effects. Here, we study In-doped SnTe nanoplates, $In_xSn_{1-x}Te$, with x ranging from 0 to 0.1 and show they superconduct. More importantly, we show that In doping reduces the bulk mobility of $In_xSn_{1-x}Te$ such that the surface states are revealed in magnetotransport despite the high bulk carrier density. This is manifested by two-dimensional linear magnetoresistance in high magnetic fields, which is independent of temperature up to 10 K. Aging experiments show that the linear magnetoresistance is sensitive to ambient conditions, further confirming its surface origin. We also show that the weak antilocalization observed in $In_xSn_{1-x}Te$ nanoplates is a bulk effect. Thus, we show that nanostructures and reducing the bulk mobility are effective strategies to reveal the surface states and test for topological superconductors.

A topological superconductor (TSC) is a superconductor possessing gapless Dirac-dispersive surface states protected by time-reversal symmetry[1,2]. Owing to this, a TSC is predicted to host Majorana fermions and provide ideal platforms to realize fault-tolerant topological quantum computers [1,2]. Superconductivity in three-dimensional (3D) topological insulators (TIs) has been shown in a Cu-intercalated $Bi_2Se_3$ bulk crystal with a critical temperature ($T_c$) of ~3.8 K[3,4]. However, whether the surface state remains intact needs further investigation[5]. Nanostructured Cu-intercalated $Bi_2Se_3$ has not demonstrated superconductivity despite demonstrations of high Cu intercalation into $Bi_2Se_3$ nanoribbons and plates [6,7]. Recently, SnTe has been rediscovered as a topological crystalline insulator (TCI) whose Dirac-dispersive surface states are protected by crystal mirror symmetry rather than time-reversal symmetry[8,9]. Superconductivity has been shown in SnTe bulk crystals doped with Indium (In) with $T_c$'s up to 4.5 K[10,11,12,13]. Angle-resolved photoemission spectroscopy[14] and zero-bias conductance peak measurement[15] indicate that the surface states remain intact after In doping, making In-doped SnTe a strong candidate of a topological superconductor.

Electrical transport studies of SnTe and In-doped SnTe are essential for both fundamental studies and potential applications, such as spintronics and quantum computing. Currently, this is seriously hindered by the high bulk carrier density of SnTe due to Sn vacancies and anti-site defects[16, 17]. Further complications arise when SnTe is doped with In to induce superconductivity as In doping further increases the already-high bulk carrier density[11]. In this regard, superconducting In-doped SnTe nanostructures are promising due to their high surface-to-volume ratios and well-defined nanoscale morphology, which can enhance the surface state effectively[18]. Here, we demonstrate that In doping in SnTe nanoplates can deliberately slow down the bulk carriers, thus revealing the surface state despite the high bulk carrier density.

We synthesized SnTe nanoplates with different In doping concentrations via the vapor-solid growth method and studied their magnetotransport at low temperatures. We demonstrate systematically that SnTe nanoplates become more diffusive with increasing In doping concentrations. $In_xSn_{1-x}Te$ nanoplates with $x > 0.05$ superconduct when cooled down to 1.7K and show linear magnetoresistance (MR) in high magnetic fields. Angle- and temperature-dependent MRs show that the observed linear MR is 2D and temperature-independent, coexisting with a 3D weak antilocalization (WAL). Thus, the linear MR is attributed to the Dirac-dispersive surface states. This is further supported by the aging study, which showed that the linear MR is sensitive to the environment while the bulk signal is stable. Our data show the surface states can be revealed in In-doped SnTe nanoplates despite the high bulk carrier density by reducing the bulk mobility and by enhancing the surface-to-volume ratio. This provides an effective route to study TCIs for fundamental studies and device applications.

**Results**

**Material preparation and characterization.**

Synthesis of In-doped SnTe nanoplates is described in Methods. Structural and chemical characterizations of In-doped SnTe nanoplates were carried out using a transmission electron microscope (TEM) (Fig. S1). Selected area electron diffraction and X-ray diffraction confirm that the crystal structure of In-doped SnTe nanoplates is the expected rock-salt crystal structure (Fig. S1 and S2). In doping concentrations were measured by energy dispersive X-ray spectroscopy (EDX) inside a TEM (Fig. S1). In was uniformly distributed in SnTe nanoplates (Fig. S1).

**Diffusive transport for bulk shown by ρ-T and ρ-B curves in $In_XSn_{1-X}Te$ nanoplates with increasing x.**

Temperature-dependent resistivity (ρ-T) and magnetoresistivity (ρ-B) of $In_xSn_{1-x}Te$ nanoplates with increasing x are shown in Fig. 1, and their transport parameters are listed in Table 1. Different transport regimes are observed with increasing x from 0 to 0.1. For pure SnTe nanoplates, metallic behavior is observed as evidenced by the monotonically decreasing ρ-T curve (Fig. 1a) and quadratic MR (Fig. 1b). For $0 < x < 0.1$, transport in the bulk becomes more diffusive with increasing x. This is shown by smaller decrease in resistivity with decreasing temperature for higher x ($\Delta\rho$(2K-300K)/ $\rho$(300K) in Table 1), which results in higher resistivity for In-doped SnTe at 1.7K than that for pure SnTe although their room temperature resistivities are similar. More diffusive bulk for higher x is clearly demonstrated by the one order of magnitude difference in bulk mobility between In-doped SnTe and pure SnTe with the same carrier density (Fig. 2a). We note that a ferroelectric phase transition is suppressed when x > 0.02, in agreement with bulk results[13]. For x > 0.05, In$_x$Sn$_{1-x}$Te nanoplates show a superconducting transition below 2 K (Fig. 1g and 1i). Resistivities drop rapidly below 1.9 and 1.88 K for Device #4 and #5, respectively. This rapid drop in resistivity was confirmed to be due to superconductivity by applying a small magnetic field of 0.3 T, which suppressed superconductivity as expected (Fig. S3b and S3d). Our result is in agreement with bulk studies[10, 11, 13], and shows increasing carrier density and decreasing bulk mobility with the In doping concentration, x (Table 1).

To study transport properties in the superconducting state, we measured magnetoresistivities of In$_x$Sn$_{1-x}$Te with a 100 μA applied current, which is much larger than critical current ($I_c$) to suppress the superconductivity (Fig. S3a and S3c). With increasing x, the amplitude change in MR decreases systematically ($\Delta\rho$(9T-0T)/ $\rho$(0T) in Table 1). WAL starts to appear in low magnetic fields starting at x of 0.02 and becomes more pronounced at higher x (Fig. 1, Fig. S4, and $\Delta\sigma$(WAL) in Table 1). These indicate that the bulk becomes more diffusive with increasing x, in agreement with the ρ-T curves and lower Hall mobility (Fig. 2a). This is due to In dopants which can induce impurity bands[19]. The slight increase of resistance with decreasing T below 100 K suggests presence of impurity bands (red arrows in Fig. 1e, 1g, and 1i). To study the impurity bands, we measured the carrier density as a function of temperature (Fig. 2b). The carrier density slowly decreases with decreasing temperature down to 100K, then quickly decreases below 100 K (red arrow, Fig. 2b). The behavior above 100 K is attributed to the semiconducting gap of SnTe while the rapid drop below 100 K is attributed to the impurity bands. The resistivity below 100 K is not proportional to $e^{-kT}$, thus the impurity bands may be delocalized[19]. Below ~ 10 K, the carrier density decreases at a different slope (blue arrow, Fig. 2b), suggesting the impurity bands freeze out below 10 K.

**3D WAL.**

WAL appears at low magnetic fields in In-doped SnTe. In 3D TIs such as $Bi_2Se_3$ and $Bi_2Te_3$, WAL has been shown to be 2D and was fitted to the 2D Hikami–Larkin–Nagaoka (HLN) equation to obtain α, which denotes the number of conduction channels[20, 21]. α of 0.5 may indicate one surface channel. For SnTe, analyzing WAL is complicated by the fact that SnTe possesses multiple surface states[22,23]. Accordingly, the value of α can range from 0.5 to 5 depending on the number of participating surface states and their interactions with the bulk state[23]. To distinguish if WAL is a surface effect in SnTe, we carry out angle-dependent magnetoresistance from 0 ° to 90 ° of Device #4 (Fig. 3b) and #5(Fig. S5a). Perpendicular magnetic field is at 0 ° while parallel magnetic field is at 90 °, normal to the applied current. The thicknesses of the two nanoplates were 346nm and 174nm, assumed to be much larger than the coherent length to avoid geometry-induced 2D effects. The WAL feature overlaps perfectly for all angles of magnetic fields, clearly indicating it is a 3D bulk effect. This is further supported by fitting WAL to the 2D HLN equation, which resulted in an unphysical value of α (Fig. S6).

**2D linear MR.**

We observe a linear MR without saturation up to 9T and with a sharp transition at ~4T in perpendicular magnetic field (red arrow in Fig. 3a and upper inset). We mark the transition as a crossover magnetic field, $B_c$. This sharp transition as well as the linear MR disappear in parallel magnetic field (lower inset of Fig. 3a). We carry out angle-dependent study to find out the nature of the linear MR (Fig. 3b). Given the cubic nature of SnTe and the isotropic bulk WAL, we assume that bulk magnetoconductance is isotropic such that the magnetoconductance taken with parallel magnetic field (90°) represents only the bulk state. Assuming that the surface and bulk are two independent conducting channels, we subtract the parallel magnetoconductance from the other magnetoconductance traces taken at various angles[20]. Figure 3c and 3d show the bulk-subtracted magnetoconductances at various angles, plotted in magnetic field and perpendicular component of the magnetic field, ($Bcos\theta$), respectively. Remarkably, the magnetoconductances at high magnetic fields overlap when plotted in the perpendicular component of the magnetic field. This shows experimentally that the linear MR is a 2D effect. The complete disappearance of the linear MR as well as the sharp transition in parallel magnetic field further supports its 2D nature.

**Temperature-dependent WAL and Linear MR.**

Temperature-dependent MRs from 1.69 K to 20 K were measured to investigate the linear MR in more detail (Fig. 4a). Above 10 K, WAL disappears (Fig. 4b) and MR starts to deviate from the linear line (Fig. 4c). Figure 4d shows this more clearly. Resistances at 0 and 9 T were measured as a function of temperature. They both increase in a similar fashion between 10 K and 100 K, caused by the impurity bands. However, their behaviors diverge below 10 K as the impurity bands begin to freeze out (Fig. 2b). The 0T resistance starts to decrease rapidly due to the emergence of WAL. In contrast, the 9T resistance continues to increase below 10 K with a different slope due to the linear MR. This indicates that below 10 K at high magnetic fields, the bulk carriers and impurity bands may not contribute to the transport signal significantly. The slope of the linear MR stays within 7% deviation below 10 K (Fig. 4c), indicating that it is temperature-independent.

**Aging effects.**

Electrical transport properties of Device #4 were measured one month after the initial measurement to study aging effects. The device was kept in nitrogen environment. The bulk WAL feature remains the same (Fig. 5a) and the Hall curve overlaps perfectly after one month, suggesting the carrier density and mobility of the bulk are kept within a difference of ~0.5% (Fig. 5b). In contrast, the linear MR changed dramatically after one month. The transition field $B_c$, at which the linear MR appears, increased significantly from 4±0.5T to 6.5±0.3T. We note that the slope of linear MR remains the same after one month (Fig. 5c).

**Discussion**

The linear MR may be due to the Dirac-dispersive surface state. The quantum limit model proposed by Abrikosov points out linear MR can appear in gapless linear-dispersive energy spectrum when only the first landau level is filled[24, 25]. In this case, both the Fermi energy $E_F$ and thermal energy $K_bT$ are much smaller than $E_1$-$E_0$ ($E_n$ means the energy of n$^{th}$ landau level)[26]. For TIs, linear MR was usually observed to coexist with Shubnikov–de Haas oscillations, suggesting the quantum limit model by Abrikosov does not apply[27, 28, 29]. A theory was developed by Wang and Lei to explain the linear MR in the presence of both gapless linear spectrum and landau level overlaps[30].

However, the linear MR can also appear due to other factors. A classical model by Parish and Littlewood shows that inhomogeneity in disordered conductors can give rise to an apparent linear MR by mixing a hall resistivity into a longitudinal resistivity[31, 32]. Our nanoplates are, however, single crystals with little defects, confirmed by selected area electron diffraction (Fig. S1). Thus the observed linear MR is not caused by inhomogeneity proposed by the classical model. Metals with open Fermi surfaces, e.g. Cu, can also have linear MR[26]. This is not our case. In TIs, linear MR can also be due to bulk factors such as cross-over between WAL and the quadratic MR[33], a geometrical factor due to the layered structure of TIs[34], and coupling between the magnetic field and the spin of bulk states[26, 35]. These cases do not apply to our In-doped SnTe nanoplates whose linear MR is from the surface states.

Here, we attribute the linear MR to the Dirac-dispersive surface states of SnTe, discussed in the models by Abrikosov[24, 25], and Wang and Lei[30]. This is supported by three observations. First, angle-dependent studies show a perfect 2D linear MR (Fig. 3) with the complete disappearance of the linear MR in the parallel magnetic field, indicating a surface effect. Second, the aging experiment shows gradual disappearance of the linear MR while the bulk remains the same (Fig. 5). This points to surface state degradation via aging, which has been observed for 3D TIs[36]. In addition, the slope of the linear MR does not change despite the aging. According to the two models, this indicates that the surface state carrier density remains the same[37, 38]. Third, the linear MR emerges only below 10 K, a temperature regime in which the bulk carriers and impurity bands are too diffusive to contribute to the transport signal (Fig. 2 and Fig.4d), further supporting the surface state origin. The slope of the linear MR is temperature-independent (Fig. 4c), in agreement with the two models[37, 38].

In summary, the surface state of topological crystalline insulator SnTe is revealed in In-doped SnTe nanoplates by reducing the bulk mobility dramatically via In doping. The 2D Linear MR at high magnetic fields is attributed to Dirac-dispersive surface states in accordance with quantum models. This is further confirmed by the aging effect, which shows that the 2D linear MR degrades over time while the bulk state remains the same. Thus we show, via electrical transport, that the surface state remains intact in the superconducting regime of In-doped SnTe nanoplates. This is crucial for fundamental studies and potential device applications based on TCIs.

**Methods**

**CVD sample growth**. In-doped SnTe nanoplates were grown using a single-zone, horizontal tube furnace with SnTe powder as a source material and InTe powder as a dopant source. VLS and VS growth of SnTe nanoplates with {100} or {111} as top and bottom surfaces has been demonstrated recently[17]. Here, we modify the growth conditions to achieve In doping. InTe was used as a dopant source because it has the same crystal structure as SnTe and its melting temperature (696 °C) is above the growth temperature. Pure In powder resulted in a doping concentration below 1 % likely due to its low melting temperature (156 °C), which is not compatible with the substrate temperatures of SnTe nanoplates (300-350 °C). Mixture of InTe and SnTe powders was placed at the center of the tube furnace and heated to 610 °C. SiOx/Si substrates coated with a ~ 10 nm thin gold film were placed at the downstream to collect In-doped SnTe nanoplates. For vapor transport, ultrapure Ar gas was used as the carrier gas. During growth, the pressure was kept at 3-5 Torr and the flow rate of the Ar gas was ~200 s.c.c.m. The growth time was 10 minutes.

**Device fabrication and electric transport measurement**. Standard e-beam lithography and lift-off techniques were used to fabricate In-doped SnTe nanoplate devices on 300 nm $SiO_2$/Si substrates. A 10/150 nm Cr/Au contact was thermally evaporated. All samples were cooled down in a Quantum Design PPMS Dynacool, which has the lowest temperature of 1.69K and the maximum magnetic field of 9T. The Hall bar configuration was used in all samples to measure the longitudinal and Hall magnetoresistances. The Hall signal was removed in the longitudinal resistance by averaging the $R_{xx}$ data over positive and negative magnetic fields.

**Acknowledgements**

Microscopy facilities used in this work were supported by the Yale Institute for Nanoscience and Quantum Engineering (YINQE). The Quantum Design Dynacool used for transport measurements was supported by MRSEC DMR 1119826. Dr. Jie Shen is supported by NSF DMR 1402600.

**Author contributions**

J. J. C. and J. S. conceived the experiment. J. S. and Y. X. carried out synthesis. J. S. characterized the synthesized nanostructures and performed transport measurements and data analysis. All authors contributed to the discussion and analysis of the data and writing of the manuscript.

**Competing financial interests**

The authors declare no competing financial interests.

**Figures**

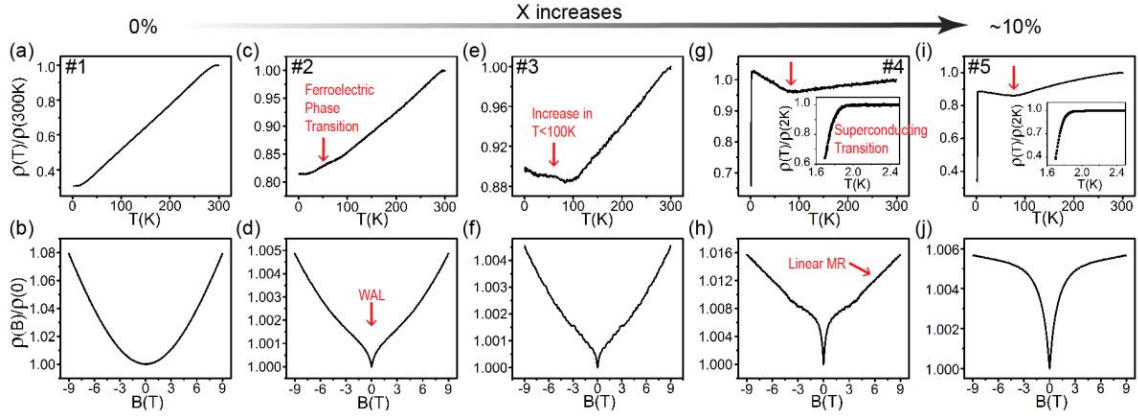

**Figure 1.** Evolution of ρ-T and ρ-B in $In_XSn_{1-X}Te$ nanoplates with increasing x. Pure SnTe (Device #1) shows a linear ρ-T curve (a) and a quadratic ρ-B curve (b), indicating metallic properties. In $In_XSn_{1-X}Te$ with x~0.02 (Device #2), a ferroelectric phase transition is observed in the ρ-T curve, indicated by the arrow (c). Also WAL starts to emerge (d). For x = 0.03 to 0.1 (Device #3, #4, #5), ρ starts to increase below 100 K (e, g, i). A superconducting sharp drop in ρ occurs below 2 K in Device #4 and #5 with x=0.05-0.1 (g, i and their insets). With increasing x, WAL becomes more obvious (d, f, h, j). Linear MR appears in Device #4 above ~4 T (h). Table 1 summarizes transport parameters of all the devices.

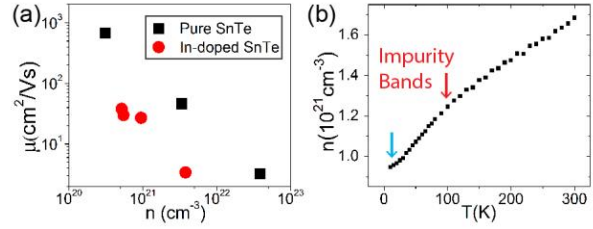

**Figure 2.** Bulk carrier densities (n) and mobilities (μ) of $In_xSn_{1-x}Te$ nanoplates. (a) shows In-doped SnTe nanoplates have lower mobilities than pure SnTe with similar carrier densities. (b) is n-T curve of Device #4, which shows three transition regions with temperature, marked by red and blue arrows in the slope changes at ~100K and ~10K, respectively. The values of mobility and carrier density are listed in Table 1. All the bulk carrier densities and mobilities are calculated from the linear Hall curve and longitudinal resistivity at zero magnetic field.

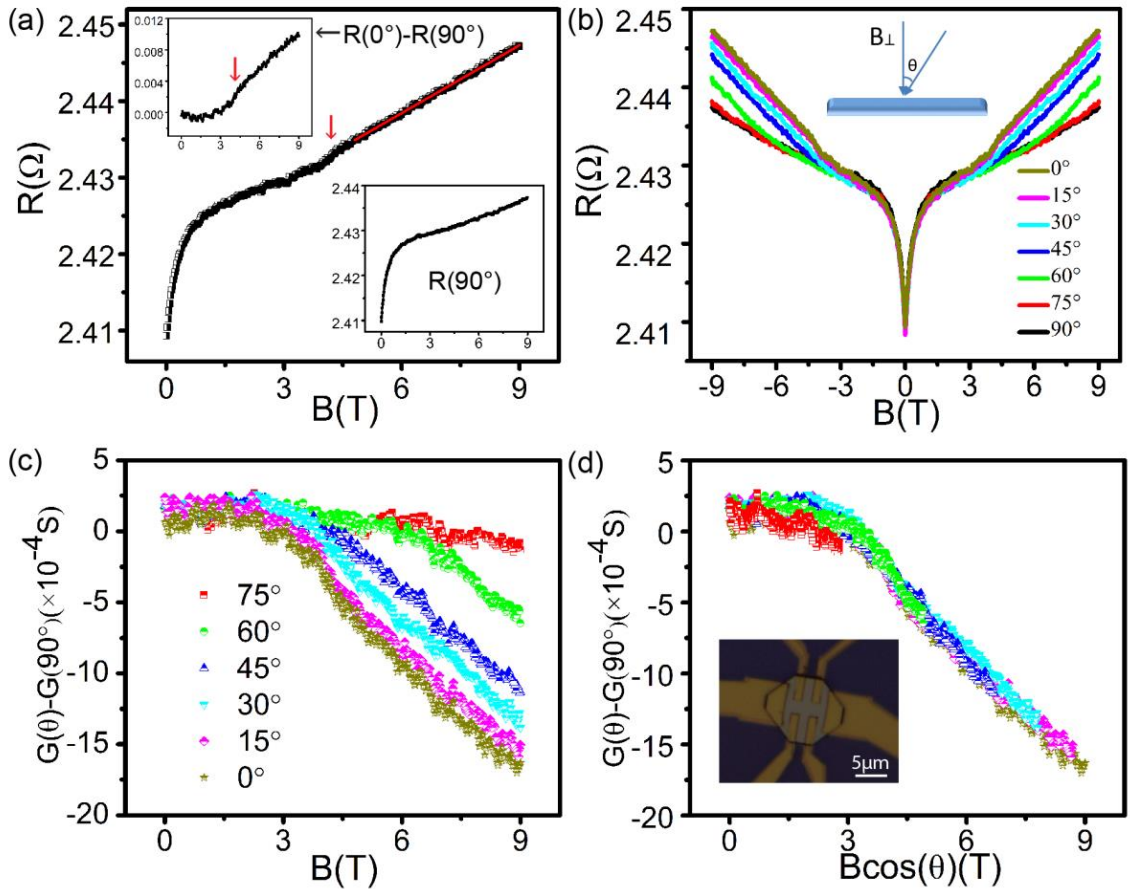

**Figure 3.** 3D WAL and 2D linear MR from angle-dependent MR of Device #4. (a) A linear MR is observed with a sharp transition, indicated by the red arrow. The red line is a linear fit to the data. Lower inset: MR measured in parallel magnetic fields does not show the sharp transition and linear property. Upper Inset: MR in perpendicular magnetic fields, after subtracting the MR in parallel magnetic fields. The transition shows up more clearly. (b) Angle-dependent MR studies show perfectly overlapping WAL, demonstrating the WAL is a 3D bulk effect. (c,d) Magnetoconductances at different angles, after subtracting the magnetoconductance taken in parallel magnetic fields, plotted in magnetic fields (c) and perpendicular component of the magnetic fields (d). The magnetoconductances overlap for all angles when plotted in perpendicualr

component of magnetic fields, showing that the linear MR is 2D. The inset in (d) is the optical image of Device #4. The direction of magnetic field is shown in the inset of (b): 0 ° means B is perpendicular to the nanoplate; 90 ° means B is parallel to the nanoplate and normal to the applied current.

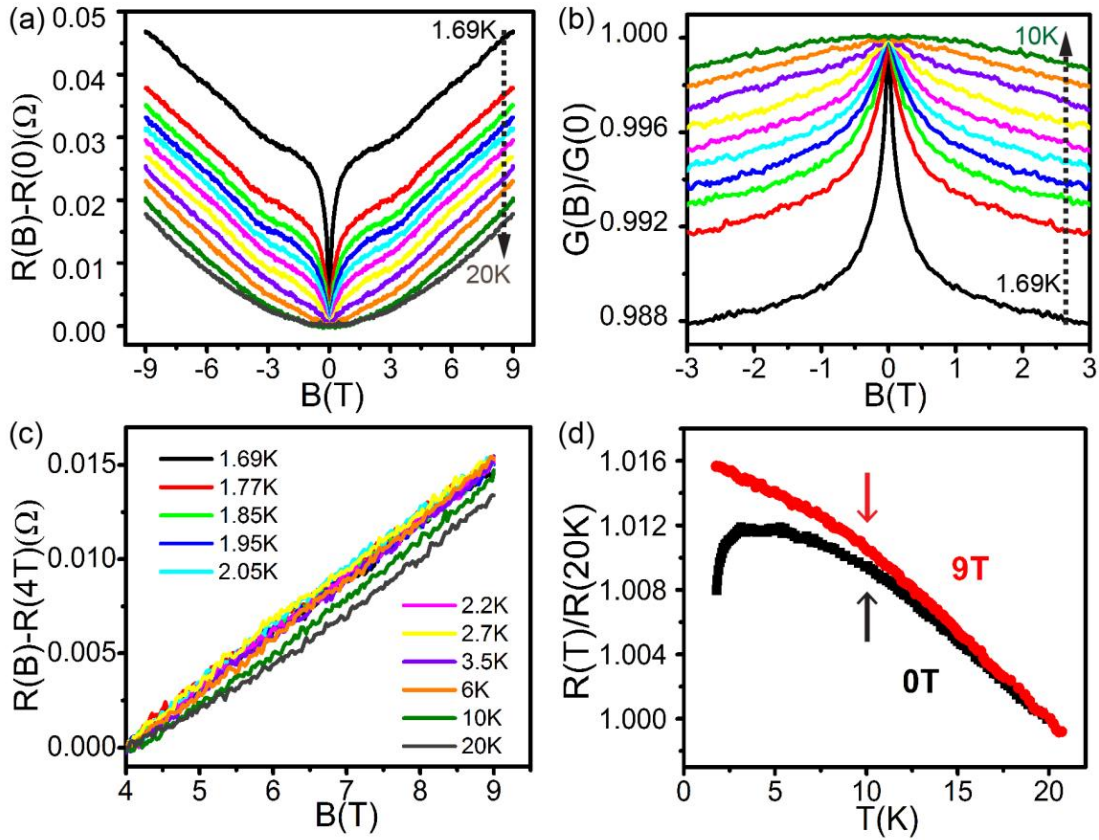

**Figure 4.** Temperature-dependent magnetoresistances of Device #4 from 1.69K to 20K. (a) R(B)-R(0) curves with increasing temperature. WAL disappears above 10K. (b) Magnetoconductances plotted up to 3 T, showing more clearly the gradual disappearance of WAL. (c) R(B)-R(4T) curves taken from 1.69 K to 20 K. The curves are linear and overlap below 10K. They begin to deviate above 10 K. (d) R-T curves taken at 0T and 9T. The transitions of their slopes are pointed out by the black and red arrows, respectively.

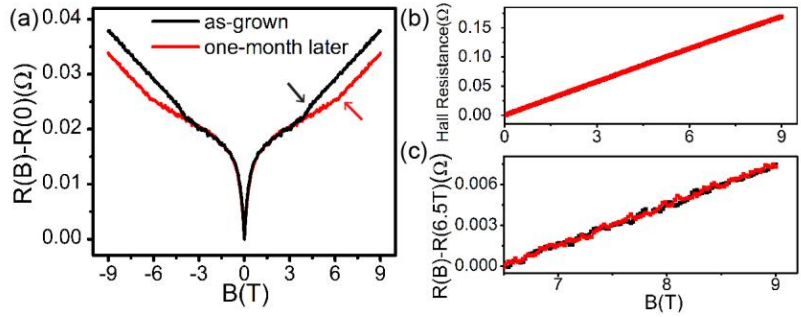

**Figure 5.** Aging experiment on Device #4. (a) MRs of Device #4 right after growth and one month later. WAL overlaps nicely to indicate bulk remains the same during aging. The linear MR, however, deviates. (b) Hall curves overlap to confirm that the bulk remains the same after one month. (c) The slope of linear MR remains the same after one month although the transition field changed. This suggests that the surface carrier density is the same.

**Tables**

|  | Pure SnTe nanoplates | | | In-doped SnTe nanoplates | | | |
|---|---|---|---|---|---|---|---|
|  | &1 | &2 | #1 | #2 | #3 | #4 | #5 |
| X (%) | 0 | 0 | 0 | ~2 | 3-6 | 5-9 | 6-10 |
| n (cm$^{-3}$) | $3.1\times10^{20}$ | $3.9\times10^{22}$ | $3.4\times10^{21}$ | $5.5\times10^{20}$ | $5.2\times10^{20}$ | $9.5\times10^{20}$ | $3.8\times10^{21}$ |
| µ (cm$^2$/Vs) | 672 | 3.2 | 46 | 30 | 38 | 27 | 3.4 |
| Δρ(2K-300K) / ρ(300K) | -72% | -70% | -69% | -18% | -11.5%, -10.4%* | -4.2%, 2.5%* | -13.9%, -11.3%* |
| Δρ(9T-0T)/ρ(0T) | 7.2% | 6.5% | 7.9% | 0.49% | 0.46% | 1.6% | 0.58% |
| Δσ(WAL) | - | - | - | 0.1% | 0.08% | 0.6% | 0.5% |

\* means the resistance increases with decreasing temperature below ~100K when doping concentration is high.

**Table 1.** Summary of transport parameters from the ρ-T and ρ-B studies of In$_X$Sn$_{1-X}$Te nanoplates with increasing x. Devices #1-5 are shown in Fig. 1. Devices &1 and &2 are two additional pure SnTe nanoplates. The carrier density and mobility are of the bulk, because they are calculated from the linear hall resistance and longitudinal resistivity at zero magnetic field.

**Supporting Information for**

# Revealing surface states in In-doped SnTe superconducting nanoplates with low bulk mobility

*Jie Shen[1,2], Yujun Xie[1,2], Judy J. Cha[1,2]\**

[1] Department of Mechanical Engineering and Materials Science, Yale University, New Haven, CT, USA

[2] Energy Sciences Institute, Yale West Campus, West Haven, CT, USA

**Corresponding Author**

\* E-mail: judy.cha@yale.edu

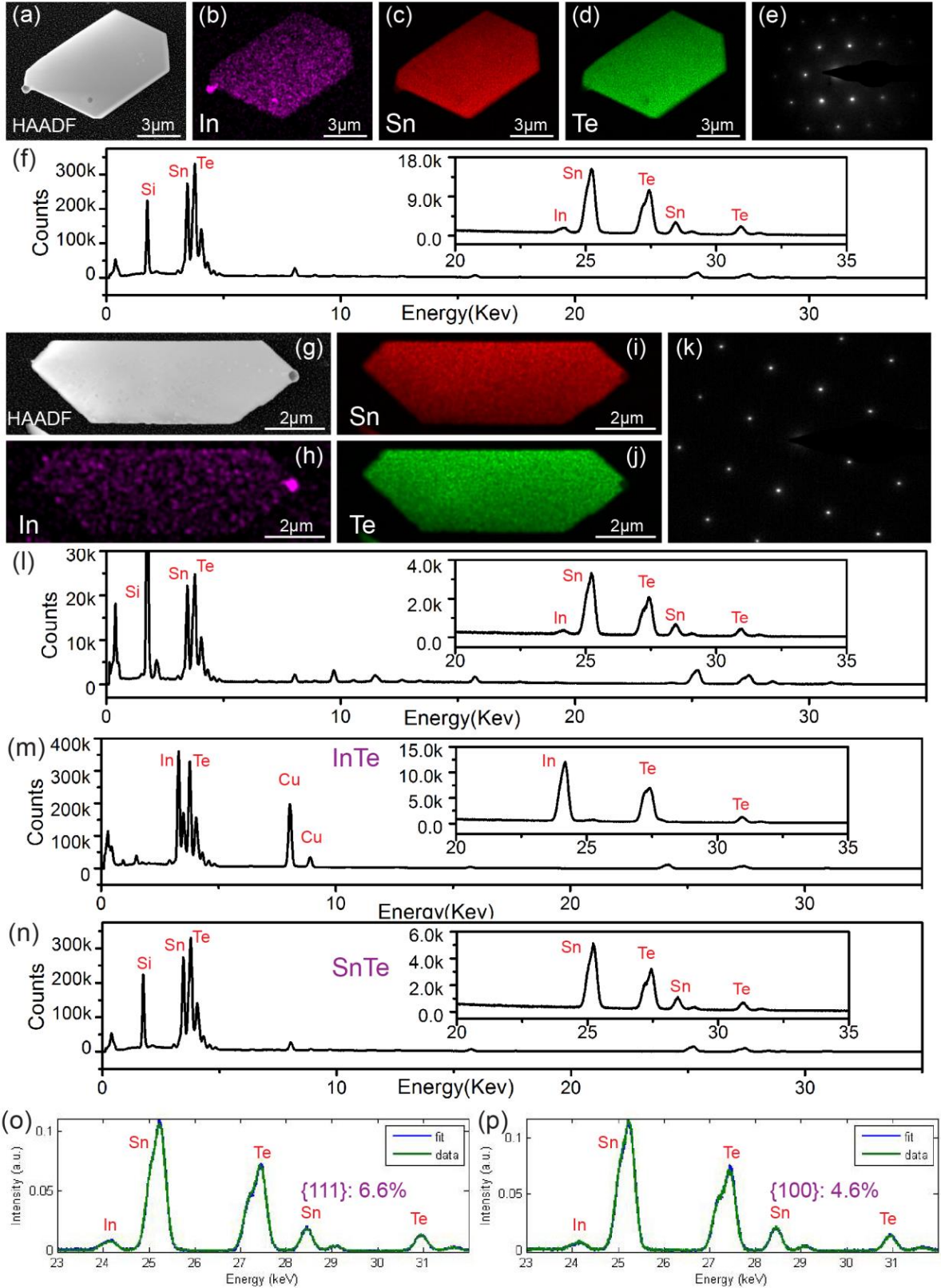

**Figure S1.** Chemical maps, EDX fit and diffraction patterns of In-doped SnTe nanoplates with {100} and {111} top surfaces. (a) High-angle annular dark field (HAADF) scanning TEM (STEM) image of a {111} nanoplate. The side facets are 60° and 120°. (b)-(d) EDX elemental maps of In, Sn and Te of the {111} nanoplate show In is uniformly distributed. (e) Electron diffraction of the {111} nanoplate shows the hexagonal symmetry, indicating that the top and bottom surfaces are (111). (f) EDX spectrum of the {111} nanoplate clearly shows the presence of In (inset). The Si peak is from the SiNx TEM grid. (g) HAADF-STEM image of a {100} nanoplate. The side facets are 45°, 90° and 135°. (h)-(j) Elemental maps of In, Sn and Te of the {100} nanoplate show the uniform distribution of In. (k) Electron diffraction of the {100} nanoplate shows the square symmetry indicating that the top and bottom surfaces are (100). (l) EDX spectrum of the {100} nanoplate clearly shows the presence of In (inset). To quantify the In doping concentration, EDX spectra of the In-doped SnTe were decomposed to reference EDX spectra of InTe and SnTe. (m) and (n) show EDX reference spectra of InTe and SnTe respectively. In, Sn, and Te peaks overlap in the energy range between 3 – 5 keV, thus the energy range was limited to 23 – 32 keV for fitting. Cu peaks are from the TEM grid. (o) and (p) are the fits for the {111} and (100) nanoplates, showing In concentrations are ~6.6% and ~4.6%, respectively. Error bars are ~ 1-2%. For electrical transport, no significant differences are observed between {111} and {100} nanoplates.

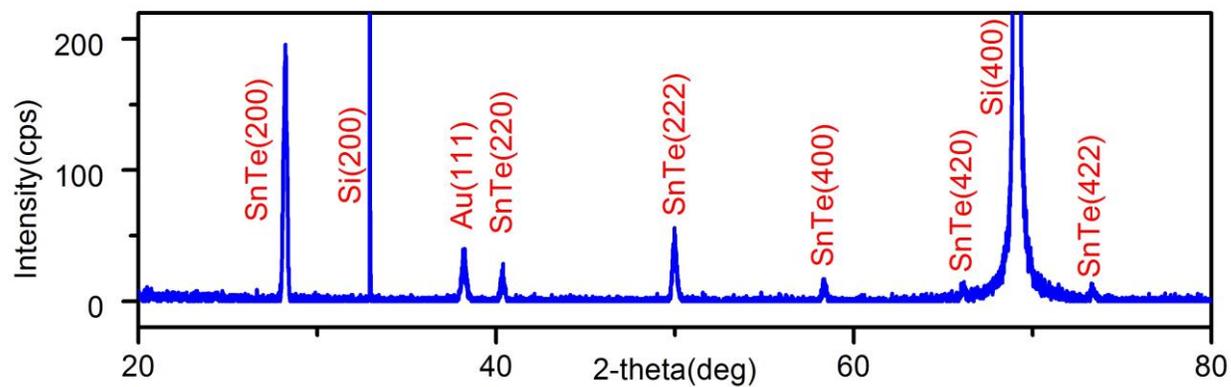

**Figure S2.** X-ray diffraction data from the In-doped SnTe growth substrate from which nanoplates were studied in the main text. Except the peaks from the substrate (Si(200) and Si(400)) and Au catalyst (Au(111)), diffraction peaks from the rock-salt SnTe are observed, indicating there is only the pure In-doped SnTe crystal in this substrate.

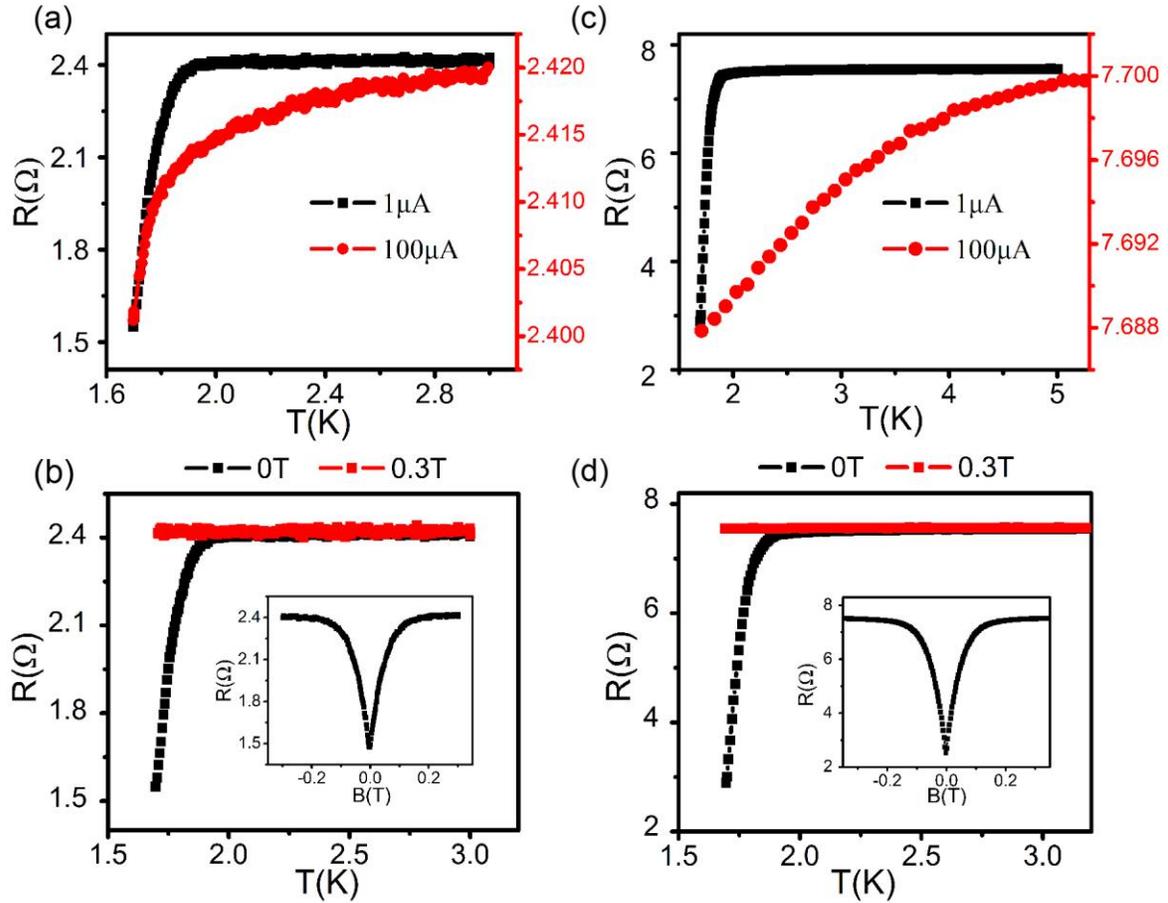

**Figure S3.** Resistance (R) versus temperature (T) at low temperatures for Device #4 (a,b) and Device #5 (c,d). (a,c) The rapid drop in resistance for Device #4 and #5 with an applied current of 1 μA is due to superconductivity. To confirm this, we applied 100 μA and measured the resistance (red traces). At 100 μA, which is expected much higher than the critical current to suppress the superconductivity, the decrease in R is very small (y axis on the right in red) and is caused by WAL. (b,d) In the presence of a magnetic field of 0.3T with 1 μA current, the sharp drop at T < 2 K disappears for Device #4 and #5, further demonstrating the sharp drop in resistance is caused by the superconducting state. The insets are the R-B curves at 1.7 K, indicating the superconductivity is suppressed when B reaches 0.2 T.

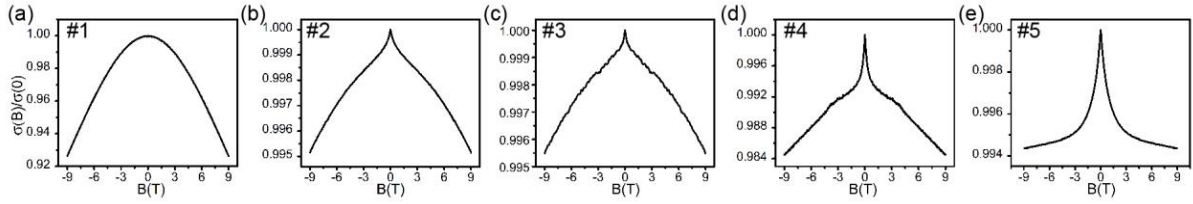

**Figure S4.** Evolution of σ-B curves in In$_X$Sn$_{1-X}$Te nanoplates with increasing x from Device #1 to Device #5. Weak antilocalization (WAL) appears in In-doped SnTe nanoplates and becomes more obvious with increasing x. All the data here are summarized in Table 1 in the main text.

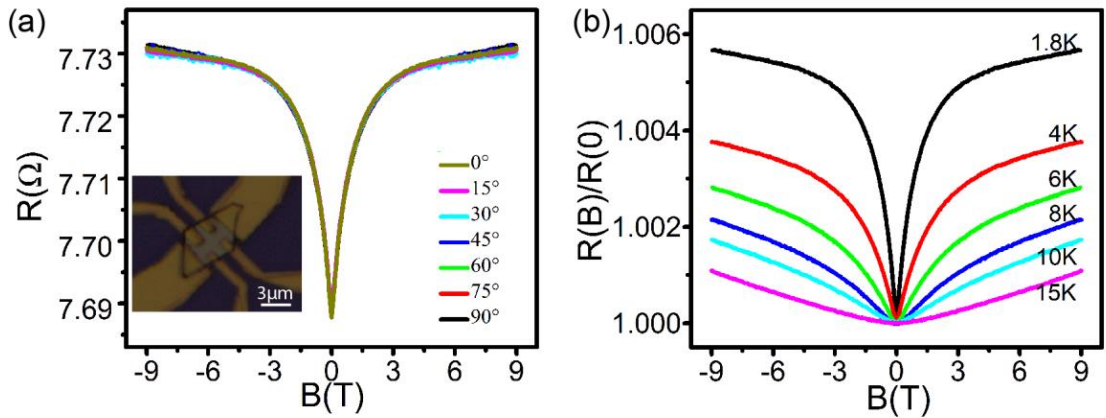

**Figure S5.** Angle- and temperature- dependent magetoresistances of Device #5. (a) The angle-dependent magnetoresistances overlap perfectly, indicating WAL is a 3D bulk effect. (b) The temperature-dependent magnetoresistances show WAL disappears above 10K.

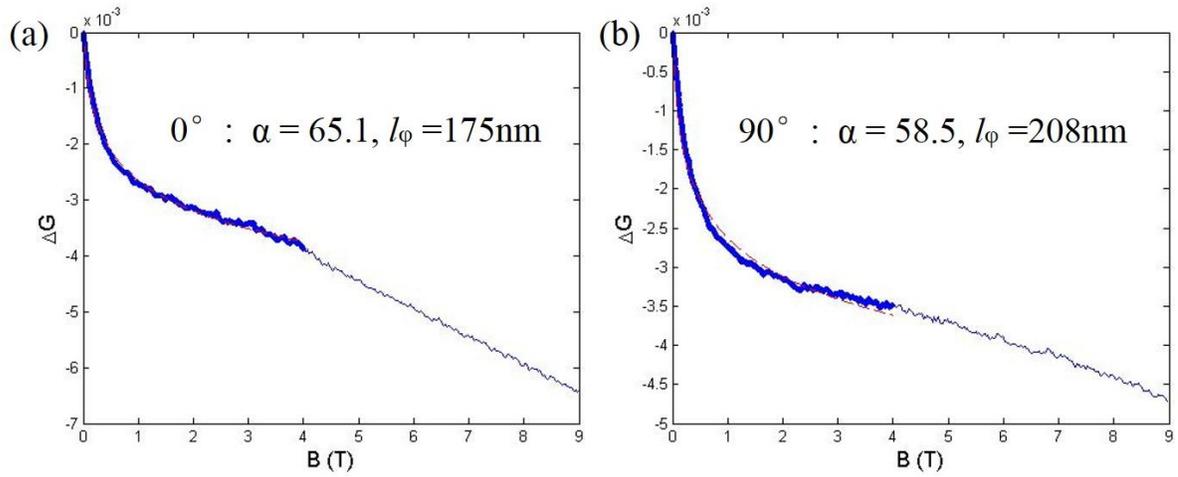

**Figure S6.** Typical WAL fitting (red dashed line) with the 2D HLN equation with different orientation of magnetic fields (0 and 90°) for Device #4. The blue region indicates the range of the fitting. The fitting is excellent but the value of $\alpha$ is too big to have any physical meaning. Also, the fitting is not robust as changing the fitting range changes the fitting parameters dramatically.